\documentclass{emulateapj}

\usepackage{wasysym,natbib}
\usepackage{epstopdf,graphicx}
\usepackage{color}

\submitted{}
\journalinfo{Submitted to ApJ}
\shorttitle{Ly Continuum of Cosmic Horseshoe}
\shortauthors{Vasei et al.}

\begin{document}

\title{ The Lyman Continuum Escape Fraction of The Cosmic Horseshoe: A Test of Indirect Estimates$^{\dagger}$ $^*$} 

\author{{\sc Kaveh Vasei\altaffilmark{1}, Brian Siana\altaffilmark{1}, Alice E. Shapley\altaffilmark{2}, Anna M. Quider\altaffilmark{3}, Anahita Alavi\altaffilmark{1}, Marc Rafelski\altaffilmark{4}, Charles C. Steidel\altaffilmark{5}, Max Pettini\altaffilmark{3}, Geraint F. Lewis\altaffilmark{6}}}

\altaffiltext{1}{Department of Physics and Astronomy, University of California Riverside, Riverside, CA 92521}
\altaffiltext{2}{Department of Astronomy, University of California Los Angeles, Los Angeles, CA 90095}
\altaffiltext{3}{Institute of Astronomy, Madingley Rd., Cambridge CB3 0HA, UK;}
\altaffiltext{4}{Goddard Space Flight Center, Code 665, Greenbelt, MD 20771}
\altaffiltext{5}{Cahill Center for Astronomy and Astrophysics, California Institute of Technology, 1216 East California Boulevard., MS 249-17, Pasadena, CA 91125}
\altaffiltext{6}{Sydney Institute for Astronomy, School of Physics, A28, The University of Sydney, NSW 2006, Australia}
\altaffiltext{${\dagger}$}{Based on observations made with the NASA/ESA Hubble Space Telescope, obtained at the Space Telescope Science Institute, which is operated by the Association of Universities for Research in Astronomy, Inc., under NASA contract NAS 5-26555.  These observations are associated with programs 11602 and 12266.}

\altaffiltext{$*$}{Some of the data presented herein were obtained at the W.M. Keck Observatory, which is operated as a scientific partnership among the California Institute of Technology, the University of California and the National Aeronautics and Space Administration. The Observatory was made possible by the generous financial support of the W.M. Keck Foundation.}
\email{svase001@ucr.edu}

\begin{abstract}

High redshift star-forming galaxies are likely responsible for the reionization of the Universe, yet direct detection of their escaping ionizing (Lyman continuum) photons has proven to be extremely challenging. In this study, we search for escaping Lyman continuum of the Cosmic Horseshoe, a gravitationally lensed, star-forming galaxy at z=2.38 with a large magnification of $\sim24$. Transmission at wavelengths of low ionization interstellar absorption lines in the rest-frame ultraviolet suggest a patchy, partially transparent interstellar medium. This makes it an ideal candidate for direct detection of the Lyman continuum.  We obtained a 10-orbit Hubble near-UV image using the WFC3/UVIS F275W filter that probes wavelengths just below the Lyman limit at the redshift of the Horseshoe in an attempt to detect escaping Lyman continuum radiation. After fully accounting for the uncertainties in the opacity of the intergalactic medium as well as accounting for the charge transfer inefficiency in the WFC3 CCDs, we find a $3 \sigma$ upper-limit for the relative escape fraction of $f_{esc,rel}<0.08$. This value is a factor of five lower than the value (0.4) predicted by the 40\% transmission in the low-ion absorption lines. We discuss the possible causes for this discrepancy and consider the implications for future attempts at both direct Lyman continuum detection as well as indirect estimates of the escape fraction.
 
\end{abstract}


\section{Introduction}
\label{sec:intro}

Star-forming galaxies are expected to be responsible for the reionization of the intergalactic hydrogen at $z>7$ \citep[e.g.,][]{Robertson15} and much of the ionizing background at $3<z<7$ \citep{Nestor13,Becker13,Becker15}. Therefore, there has been much interest in quantifying the fraction of ionizing photons that escapes from star-forming galaxies (the ``escape fraction,'' $f_{esc}$). 

Recent measurements of the UV luminosity functions of galaxies \citep[e.g.][]{Oesch13,Alavi14,Bouwens15a,Atek15} have shown that galaxies can provide enough ionizing photons by $z \sim 6$ if the luminosity function is extrapolated to luminosities beyond our current detection limits and if the escape fraction is high  \citep[$\sim 0.2$ or more][]{Bouwens12,Robertson13,Robertson15, Bouwens15b}. However, the escape fraction is not well constrained, nor is the mechanism allowing for leakage of ionizing (Lyman continuum, LyC) photons. 
Unfortunately direct detection of the ionizing photons from galaxies at the epoch of reionization or soon thereafter is not feasible due to the high opacity of the intergalactic medium (IGM) at $z \gtrsim 4$ \citep[][]{Fan06,Prochaska09}. Hence, over the last two decades, many attempts have been made to detect escaping LyC from various types of star-forming galaxies at $z \lesssim 4$ \citep[e.g.][]{Leitherer95,Steidel01,Grimes07,Grimes09,Siana07,Siana10,Siana15,Cowie09,Iwata09,Bridge10,Vanzella10,Boutsia11,Nestor11,Nestor13,Mostardi13,Guaita16}.

At $z<2$, despite higher IGM transmission and lower foreground contamination rates, only 4 LyC emitters have been identified \citep[][]{Leitet11,Leitet13,Borthakur14,Izotov16}, with escape fractions less than $4\%$ in three of the galaxies and $\sim8\%$ in the other. 
At $2<z<4$, ground-based studies have yielded many LyC-emitting candidates. However, after careful re-examination of many of them with higher resolution HST images,  only three robust detections have been confirmed \citep[][]{Vanzella12,Vanzella15,DeBarros15,Mostardi15,Vanzella16}.     

Due to the limited success in LyC direct detection, many studies have tried to indirectly determine the escape fraction \citep[e.g.][]{Heckman11, Jones13,Borthakur14,Alexandroff15,Erb15} based on the assumption of a ``picket fence'' model of the interstellar medium, where the foreground absorbing gas is assumed to be patchy, where parts of the galaxy are covered by opaque gas clouds and others are not. Thus, $f_{esc}$ is not dictated by a single column density of foreground gas, but is instead equal to the fraction of the galaxy not covered by opaque, {\it neutral} gas clouds. In this scenario, the covering fraction can be estimated using strong absorption lines from low ions (assumed to be cospatial with neutral hydrogen), where the transmitted UV continuum in the saturated core of an absorption line (which would normally be zero for complete coverage) should give a good indication of the fraction of the UV disk that is transparent and, thus, the LyC escape fraction.

The majority of ionizing photons are emitted by O-stars that are formed in dense molecular clouds. Therefore, mechanisms to expose these stars must occur on relatively short timescales (within O-star lifetimes, $< 10$ Myr). This might be achieved either by SNe feedback \citep[e.g.][]{Dove00,Fujita03,Ma15}, interactions with other galaxies \citep[][]{Gnedin08}, or runaway massive stars \citep[e.g.,][]{Conroy12}. Although the details of the mechanisms by which LyC photons can escape the host galaxy are not well established, they can be distinguished on the basis of the morphological distribution of the escaping LyC in spatially resolved, directly-detected LyC emitters. 

In this study we investigate the ionizing emission of J1148+1930, the ``{\it Cosmic Horseshoe}" \citep[]{Belokurov07}. The Horseshoe is a star-forming galaxy at $z= 2.38$ gravitationally lensed into an almost complete Einstein ring of 5$''$ radius by a massive galaxy at $z = 0.44$. The magnification factor is $24 \pm 2$ \citep[][]{Dye08}. The high resolution, rest-frame UV spectrum of the Horseshoe has multiple, resolved interstellar absorption lines of low ions with depths of $\sim60$\% of the continuum flux density \citep[][]{Quider09}. As will be discussed in Section~\ref{sec:escape} and also in the Appendix, based upon these lines the picket fence model for the Horseshoe implies a rather high $f_{esc}$. 
  
Thus, the Horseshoe is an excellent candidate from which to directly detect and study escaping LyC. In addition, its redshift is ideal for the study of LyC as the {\it Hubble Space Telescope} WFC3/UVIS F275W filter only has transmission at wavelengths just short of rest-frame 912\AA, allowing a direct measurement of LyC at wavelengths very close to the Lyman limit, where the IGM opacity is at a minimum. Moreover its large magnification provides a rare opportunity to study the Lyman continuum escape fraction with higher sensitivity and higher spatial resolution, which may allow us to distinguish between several possible LyC leakage mechanisms.

In Section 2, we define our strategy to measure the escape fraction. In Section 3, we present the data. In section 4, we outline the method to overcome the detector degradations and artifacts that limit our sensitivity. Finally, in Section 5, we present the results and discuss their importance for future investigations of escaping LyC. 
 
 Throughout the paper, we adopt a cosmology with $\Omega_{M}=0.3, \Omega_{\Lambda}=0.7$ and $H_0=70$ km s$^{-1}$ Mpc$^{-1}$. The flux densities are all in $f_{\nu}$, i.e. are given in erg s$^{-1}$ cm$^{-2}$ Hz$^{-1}$, and the magnitudes are in the AB system.

\section{Escape Fraction Definition}
There are two widely used definitions of the escape fraction. First is the absolute escape fraction, $f_{esc,abs}$, defined as the fraction of emitted LyC photons that escapes into the IGM. This definition is convenient for theoretical models where one can translate star formation rates into intrinsic LyC flux, but difficult to measure observationally. To overcome the uncertainties of using the absolute escape fraction,  \citet[]{Steidel01} defined the relative escape fraction, $f_{esc,rel}$, as the fraction of escaping Lyman continuum photons divided by the fraction of escaping photons at rest-frame 1500 \AA:  
\begin{equation}
{f_{esc,rel}=\frac{{{\left( F_{out}/F_{stel}\right)}}_{LyC}}{{\left( F_{out}/F_{stel}\right)}_{1500}}}=\frac{f_{esc,abs}}{10^{-0.4A_{1500}}} 
\end{equation}
where $F_{out}$ is the flux that escapes the host galaxy into the IGM, thereby contributing to the ionizing background and $F_{stel}$ is the total intrinsic flux produced in the galaxy. $A_{1500}$ is the dust attenuation, in magnitudes, at 1500\AA. For a Calzetti reddening law \citep[]{Calzetti97} we would have \(A_{1500}=10.33 E(B-V)\). Rearranging the equation gives the commonly used equation for relative escape fraction \citep[][]{Siana07}: 
\begin{equation}
\newcommand{\RNum}[1]{\uppercase\expandafter{\romannumeral #1\relax}}
{f_{esc,rel}=\frac{{\left( F_{1500}/F_{LyC}\right)}_{stel}}{{\left( F_{1500}/F_{LyC}\right)}_{obs}}\times e^{{\tau}_{H_{\RNum{1},IGM}}}}\label{eq:fesc}
\end{equation}  
where \({\tau}_{H_{I,IGM}}\) is the optical depth of the Lyman line and continuum absorption of the neutral intergalactic hydrogen along the line of sight (hereafter LoS), \({(F_{1500}/F_{LyC})}_{stel}\) is the intrinsic flux decrement across the Lyman break, and \({(F_{1500}/F_{LyC})}_{obs}\) is the corresponding observed ratio.
Eq.~\ref{eq:fesc} is useful as it does not  require knowledge of dust attenuation and the 1500 \AA\ flux density of high redshift galaxies can be easily measured. 

\begin{figure*}
\plottwo{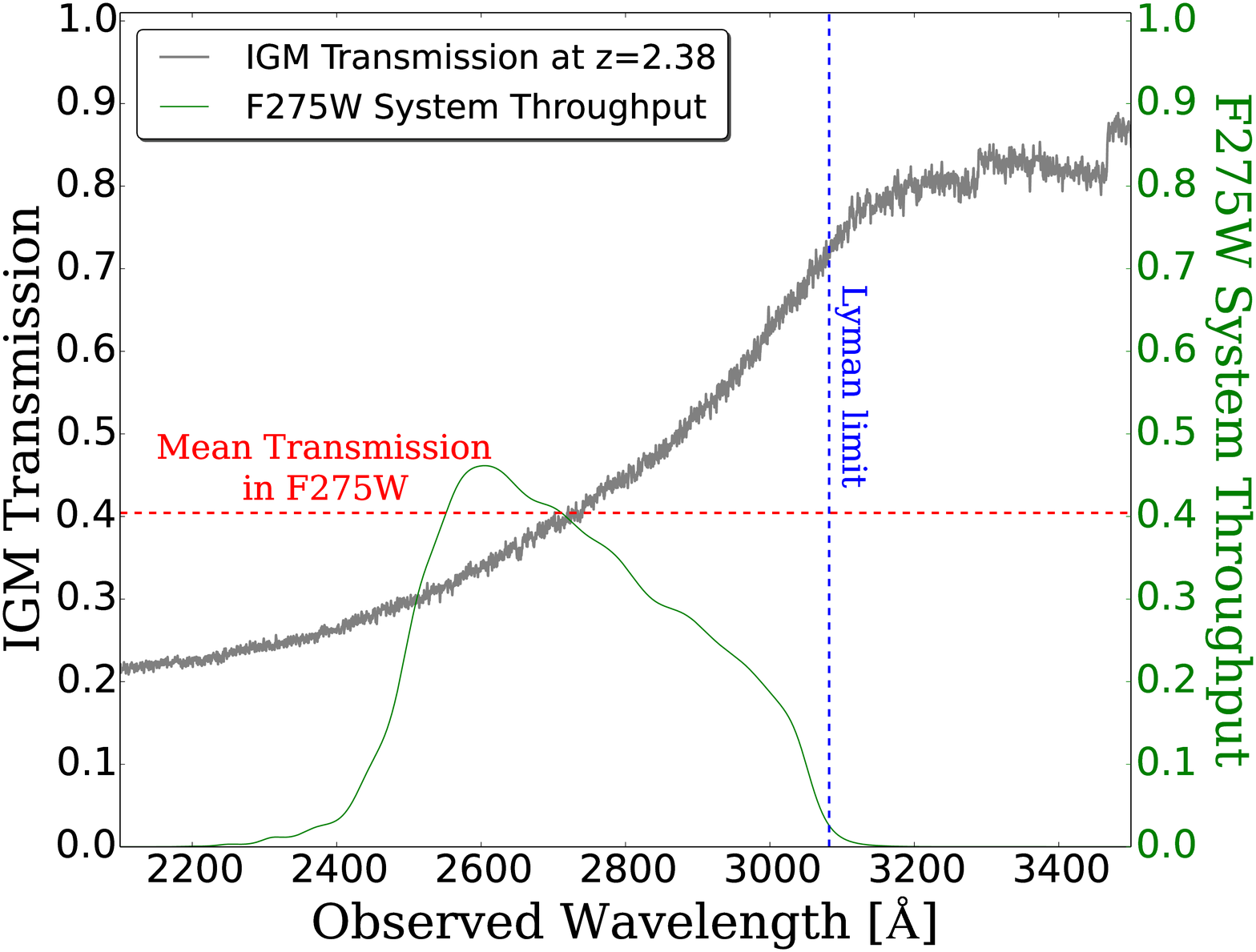}{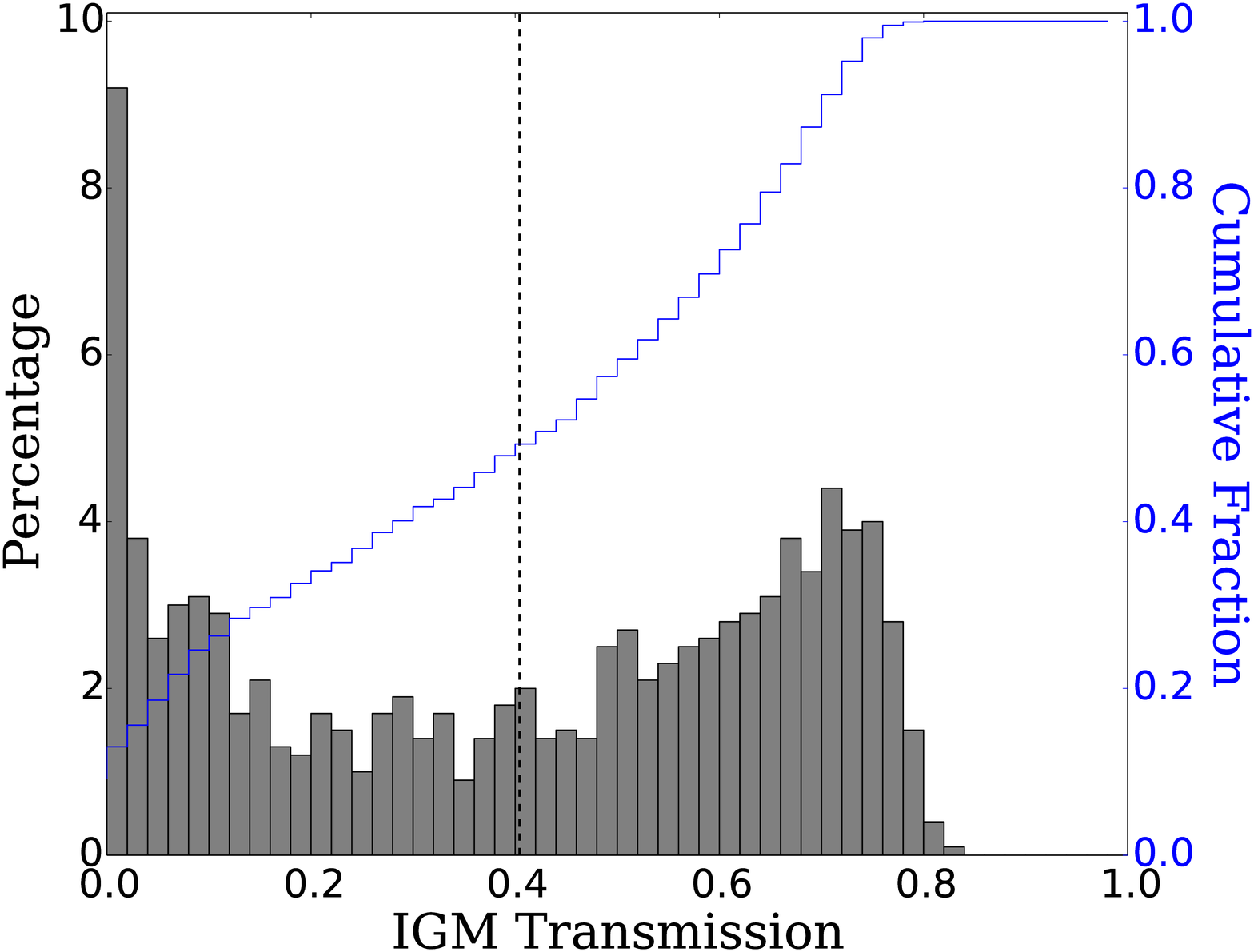}
\caption{Left: The average transmission (from 1000 simulated LoS) of Lyman continuum photons through the IGM from galaxies at $z=2.38$. The horizontal, red, dashed line indicates the F275W system throughput-weighted average IGM transmission (0.40). Right: A histogram of the F275W filter transmission-weighted IGM transmission of the 1000 simulated LoS. The vertical dashed line shows the same average transmission (0.4) through the filter and the blue line shows the cumulative fraction up to any given IGM transmission. We note that only $\sim$20\% of the LoS have transmission below 8\%.}
\label{fig:IGMT}
\end{figure*}

In order to assess the escape fraction of a galaxy we have to estimate those three values. The intrinsic flux ratio depends on age, star formation history, initial mass function (IMF) and metallicity of the stellar populations. In \citet[][their Fig 1]{Siana07} this ratio is calculated at both $\lambda_{rest}=700$ \AA\ and $\lambda_{rest}=900$ \AA\ for an instantaneous burst and continuous star formation, and using stellar population synthesis models from both \citet{Bruzual03} and \citet[][]{Leitherer99}. 

Based on the characteristics of the Horseshoe, we assumed continuous star formation which has been proceeding for the last 100 Myr \citep[][]{Quider09}. Based on this continuous star formation history, and the fact that the central wavelength of the F275W filter is at $\lambda_{rest}\sim800$ \AA\, we estimate the intrinsic flux ratio to be $(F_{1500}/F_{LyC})_{stel}\sim7$.

To estimate the average transmission of the IGM, we ran a Monte Carlo simulation \citep[see ][for details]{Siana07} using the known distributions of H{\sc i} absorber column densities as a function of redshift \citep[][]{Janknecht06,Rao06,Ribaudo11,OMeara13} as summarized in Table 2 of \citet[][]{Alavi14}.  
We simulate 1000 lines of sights (hereafter LoS) through the IGM to $z=2.38$.  Each LoS gives the transmission as a function of wavelength, from which we determine average transmission through the F275W filter as illustrated by the left panel of Fig.~\ref{fig:IGMT}.  
The right panel of Fig.~\ref{fig:IGMT} shows the distribution of IGM transmission through the F275W filter for all 1000 LoS, demonstrating significant variation. In fact, the mean value of exp($-\tau_{IGM})=0.4$ is not a common value, as the distribution is bimodal. This can be converted to the distribution of escape fraction which will give the probability of each escape fraction limit. Also \citet[][]{Rudie13}, showed considering the higher incidence of the absorbers in the circumgalactic medium (CGM) can reduce the average transmission of CGM +IGM by a factor of $\sim10$\% compared to the average transmission of only IGM, yet they still have a higher mean transmission estimate.

With estimates of the intrinsic Lyman break amplitude, $(F_{1500}/F_{LyC})_{stel}\sim7$, and the mean IGM transmission, exp($-\tau_{IGM})=0.4$, we need only to measure the flux density ratio, \({(F_{1500}/F_{LyC})}_{obs}\), to determine the relative escape fraction.
\\

\section{Observations and Data Reduction}

\subsection{Observations}
For our analysis, we have used two images from the {\it Hubble Space Telescope} Wide Field Camera 3 (WFC3) UV channel (UVIS). First, we used a one-orbit (2412 s) image in the F606W filter from the Hubble Program ID 11602 to measure the non-ionizing UV flux density at $\lambda_{rest} \sim1800$ \AA. 
 
Because the rest-UV SED is nearly flat in $f_{\nu}$ \citep[$g-i=0.04$ AB, ][]{Belokurov07}, we assume that the flux density at 1500 \AA\ (typically used for escape fraction measurements) is the same as the flux density at 1800 \AA.
 
\begin{figure*}[t]
\centering
\includegraphics[angle=0,width=2.0\columnwidth]{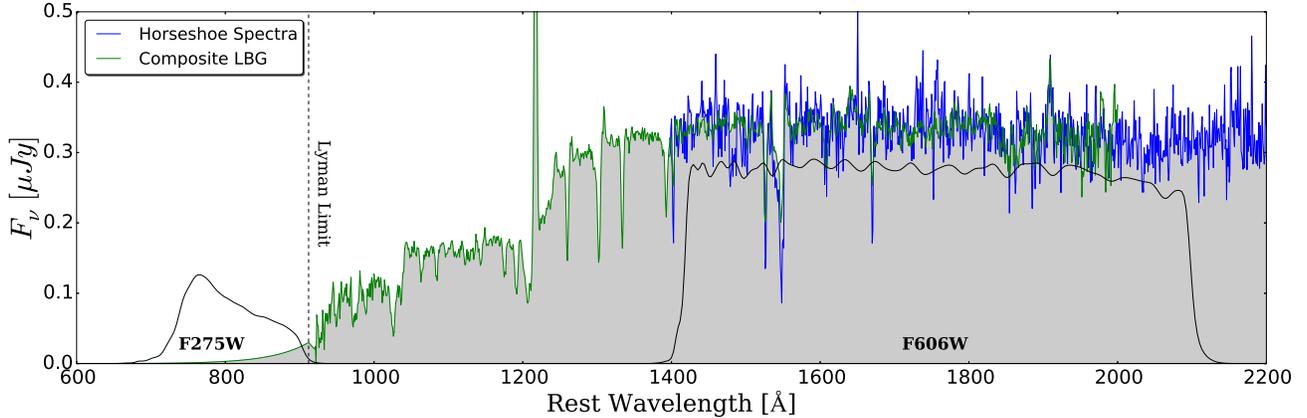}
\caption{The green line is the composite spectrum of $z\sim3$ LBGs from \citet{Shapley03}, shifted to the Horseshoe's redshift, $z = 2.38$. The blue line is the smoothed spectrum of the Horseshoe from \citet{Quider09}. The spectrum below the Lyman limit is just an illustrative extrapolation assuming a constant $f_{\nu}$ corrected for the mean IGM transmission from $z=2.38$ at each wavelength, i.e. $exp(-\tau_{ _{IGM}})$. Also plotted are the total system throughput for the two WFC3/UVIS F275W and F606W filters, which sample, respectively, the ionizing and non-ionizing UV continua.}
\label{fig:filters}
\end{figure*}

We obtained a 10-orbit image in the F275W filter to sample the rest-frame LyC, (transmission-weighted wavelength of 2704 \AA , or $800$ \AA\ in the rest frame of the Horseshoe). Fig.~\ref{fig:filters} shows these filter transmission curves with a typical Lyman break galaxy, LBG spectrum at $z=2.38$ and the smoothed spectrum of \citet{Quider09}.  

Below 4000 \AA , read noise is the main source of noise in our UVIS imaging, although, for more recent UVIS observations, the dark current has become more important. Hence, to minimize the number of readouts, the F275W exposure times were half an orbit in duration (1404 s), with a total of 20 exposures and total exposure time of  28080 s. 

The UVIS CCDs suffer from significant charge transfer inefficiency (CTI) when both the background and the targets are faint \citep[][]{MacKenty12}. As there is effectively no sky background in F275W ($\sim0.35$ $e^{-}$ per pixel per half orbit exposure), the dark current ($\sim1.5$ $e^{-}$ per pixel per half orbit exposure) dominates the total background but is very low. 
In addition, the expected LyC signal is very faint. Therefore, CTI is of particular concern for these observations and we have to understand its effects on our measurements. 
In order to minimize the number of pixels traversed by the electrons upon readout, and therefore reduce the charge transfer inefficiency, we intentionally placed the Horseshoe closer to the readout edge of the detector, with its center $\sim40\%$ of the CCD width from the read-out edge. 
In F606W, both the high signal and high background minimize CTI, so it will not significantly affect our measurements in that filter.  
\subsection{Data Reduction}

Part of the WFC3/UVIS calibration process is the subtraction of a dark reference file to correct dark current structure and flag hot pixels that can cause significant artifacts in the images. The current standard processing of the dark calibration is insufficient for the WFC3/UVIS data of this program due to the radiation damage causing poor charge transfer efficiency (CTE), and the low background levels in the F275W filter. The STScI dark frames have the following inadequacies: First, the STScI process uses an outdated definition of a hot pixel that has not been updated to account for CTE degradation, resulting in unmasked warm-to-hot pixels remaining. Second, the standard processing uses the median value of the average darks as the value of all pixels in the dark frame. This median dark file is not suitable for the low background of the NUV, because it leaves a low-level gradient and a blotchy pattern in the dark that is not subtracted. Lastly, the STScI darks are not corrected for CTE, furthering the improper hot pixel masking, and contributing to incorrect median levels in the STScI darks.

To solve these issues, custom CTE-corrected superdarks are created in a two step process as detailed in \citet[][]{Rafelski15}. First, all darks from a 4 day window at the same cadence as the STScI darks are used to create a superdark, where the background is modeled with a third order polynomial to remove the background gradient temporarily to find the hot pixels with a uniform updated threshold level. This step is necessary due to the large number of new hot pixels per day, and the drastic change in hot pixels after each anneal, where the CCD is warmed up for several hours to reduce the number of hot pixels caused by radiation damage. Then, all darks from a single anneal cycle are averaged together avoiding the hot pixels from each 4 day window, to determine the actual dark level for each good pixel. This averaged background is then used in conjunction with the hot pixel map from the first step to create a new superdark which is used for calibrating the science exposures. These new darks properly flag the hot pixels, remove the background gradient, and significantly reduce the blotchy pattern in the science exposures.

We first CTE-corrected and dark subtracted all of the flat fielded images. We then used the AstroDrizzle package \citep{Gonzaga12}, provided by the Space Telescope Science Institute, to combine the dithered images into a final image with a 0.04$''$ pixel scale. AstroDrizzle outputs inverse variance maps which were used to determine the expected Poisson uncertainties in each pixel. The alignment of individual exposures was accurate to $<4$ mas in both filters.  After combining all images, we subtracted a slight, residual background gradient in the read-out direction by fitting a third-order polynomial to both CCDs in order to guarantee a uniform background around the Horseshoe. 

\begin{figure}[b]
\includegraphics[angle=0,trim=1cm 0cm 0.5cm 0.5cm,clip=true,width=\columnwidth]{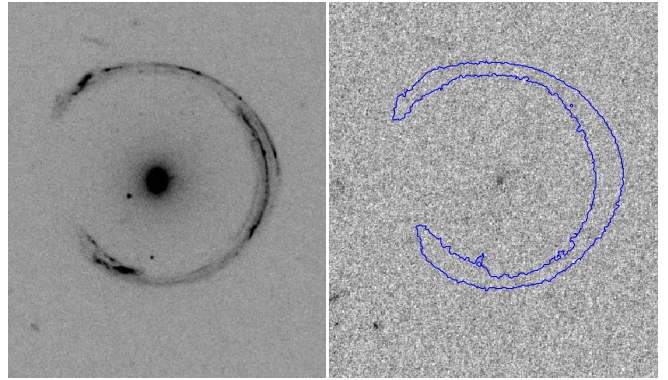}
\epsscale{.50}

\caption{Left: The 1-orbit F606W image sampling the non-ionizing flux at $\lambda_{rest}\sim1800$ \AA.  Right: The 10-orbit F275W image, with our largest aperture depicted in blue (the ``L'' aperture, defined by a low surface brightness isophote in the F606W image).}  
\label{fig:horseshoe} 
\end{figure}

The reduced images in both filters are shown in Fig.~\ref{fig:horseshoe}.   
To determine the image depths, we corrected the pixel rms for the adjacent pixel correlations based on \citet[]{Casertano00} (in our case, multiplied by 1.5) to estimate sigma per pixel. The derived $5\sigma$ image depths in an aperture with a 0.4$''$ radius are 27.34 and 26.79 AB magnitude for the F606W and F275W images, respectively. We note that this depth does not reflect the CTI that will be discussed later and affects the calculated upper limit on $f_{esc}$.     
\begin{figure}[b]
\centering
\includegraphics[angle=0,width=0.32\columnwidth]{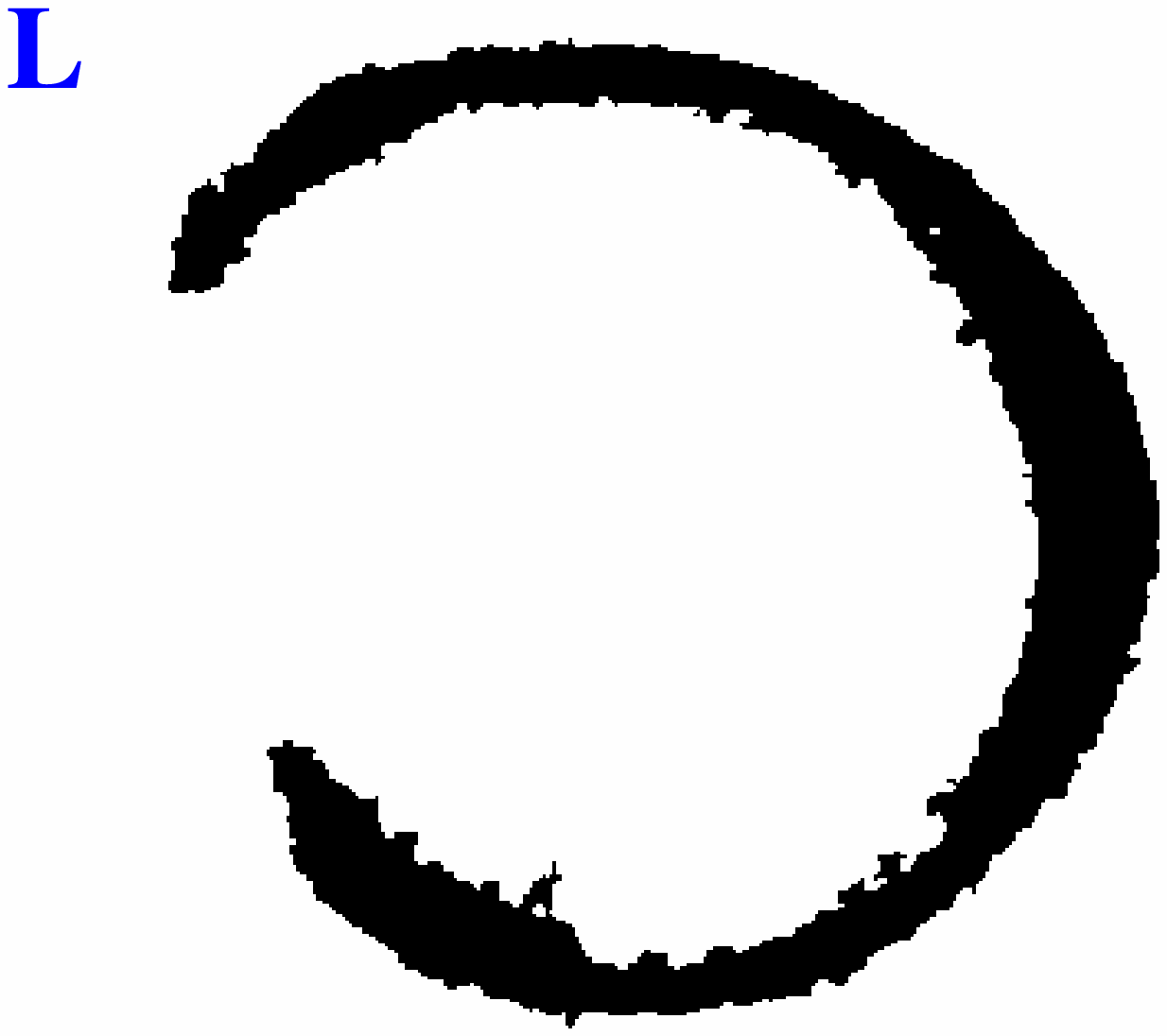}
\includegraphics[angle=0,width=0.32\columnwidth]{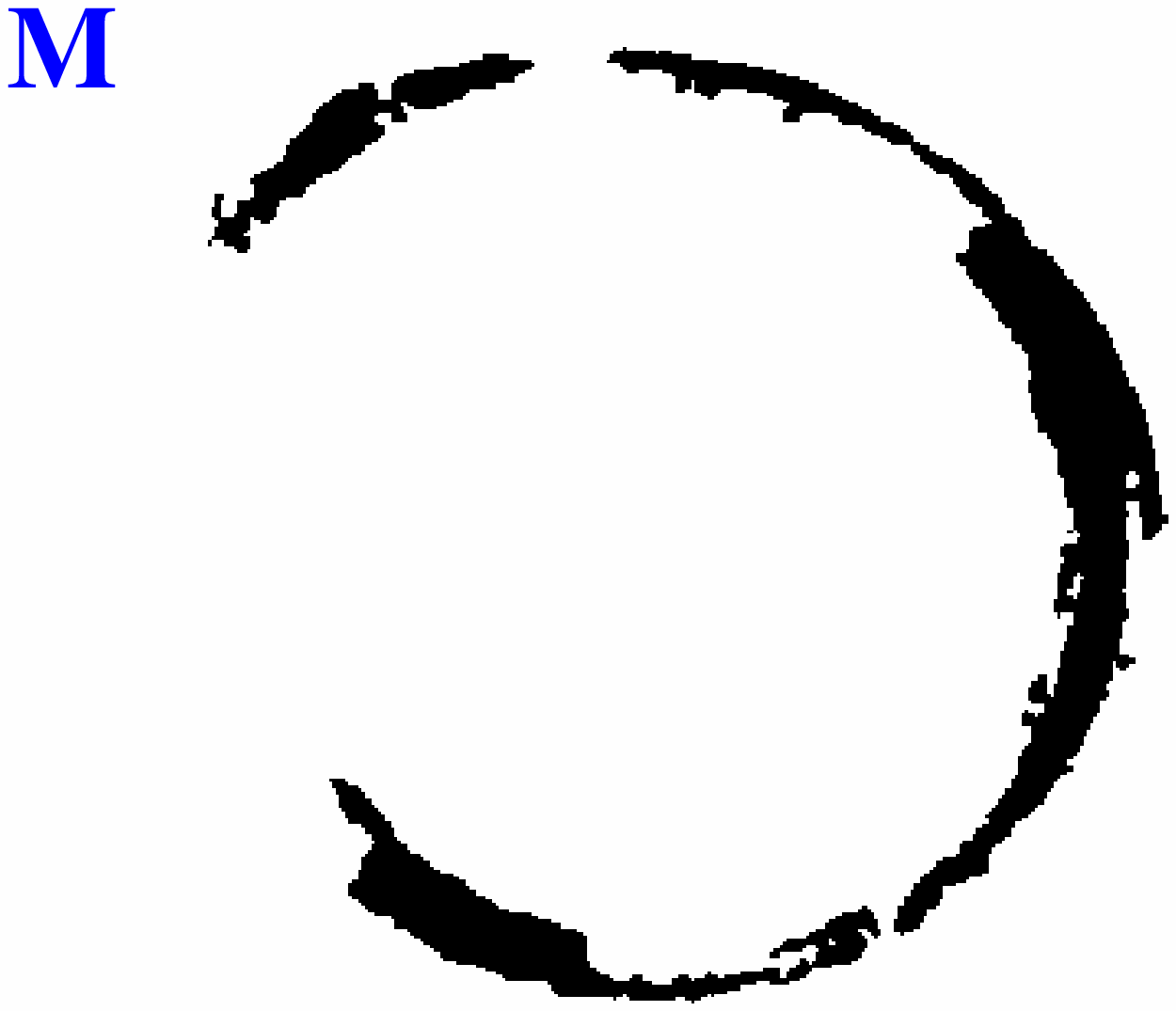}
\includegraphics[angle=0,width=0.32\columnwidth]{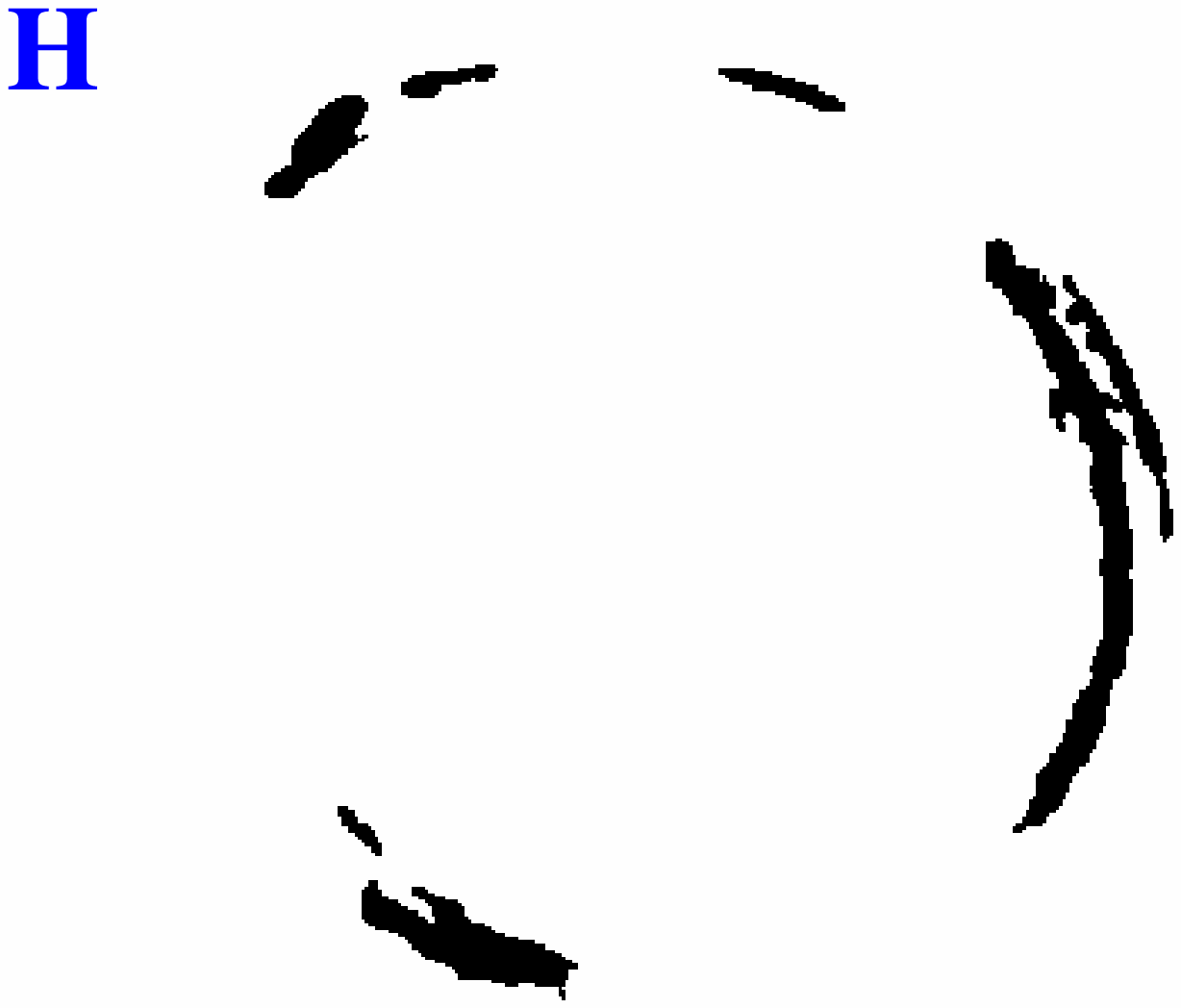}
\caption{ The ``L'' ,``M'' and ``H''  apertures defined by low, medium and high surface brightness isophotes in the F606W image (1.3, 3 and 5 $\sigma$ above the background, respectively). }
\label{fig:aps}
\end{figure}
\subsection {Photometry}

We searched for F275W (LyC) flux in a number of apertures defined by isophotes of varying F606W (non-ionizing UV) surface brightness, because it is possible that LyC is only escaping from certain regions (like those of high star formation surface density, for example),. We used three apertures with boundaries defined by low (L), medium (M), and high (H) surface brightness isophotes, corresponding to  $1.3\sigma$, $3\sigma$, and $5\sigma$ above the background in F606W. These apertures constitute a range from a near-total flux aperture up to a high surface brightness aperture and are displayed in Fig.~\ref{fig:aps}. No signal is detected through the F275W filter in any of the apertures.The F275W $S/N$ in these apertures varies between -1.0 to +1.1, and is therefore not a significant detection.

We can use the depths of the F275W image to place upper limits on the LyC escape fraction.  However, first, we must assess the effect (if any) of CTI as it may move some of the expected flux out of our aperture, erasing some of the signal. \citet[][]{Teplitz13} has shown cases where the signal can be lost completely in smaller apertures. Unfortunately, at the time the data were taken the ``post flash'' option \citep[]{Biretta13} to add background to the image was not available to mitigate CTI, and pixel-based empirical corrections of CTI are impossible in reconstructing images if no signal has remained. Therefore, our images in the F275W filter will be affected by CTI.

\begin{figure*}
\centering
\includegraphics[angle=0,scale=.24]{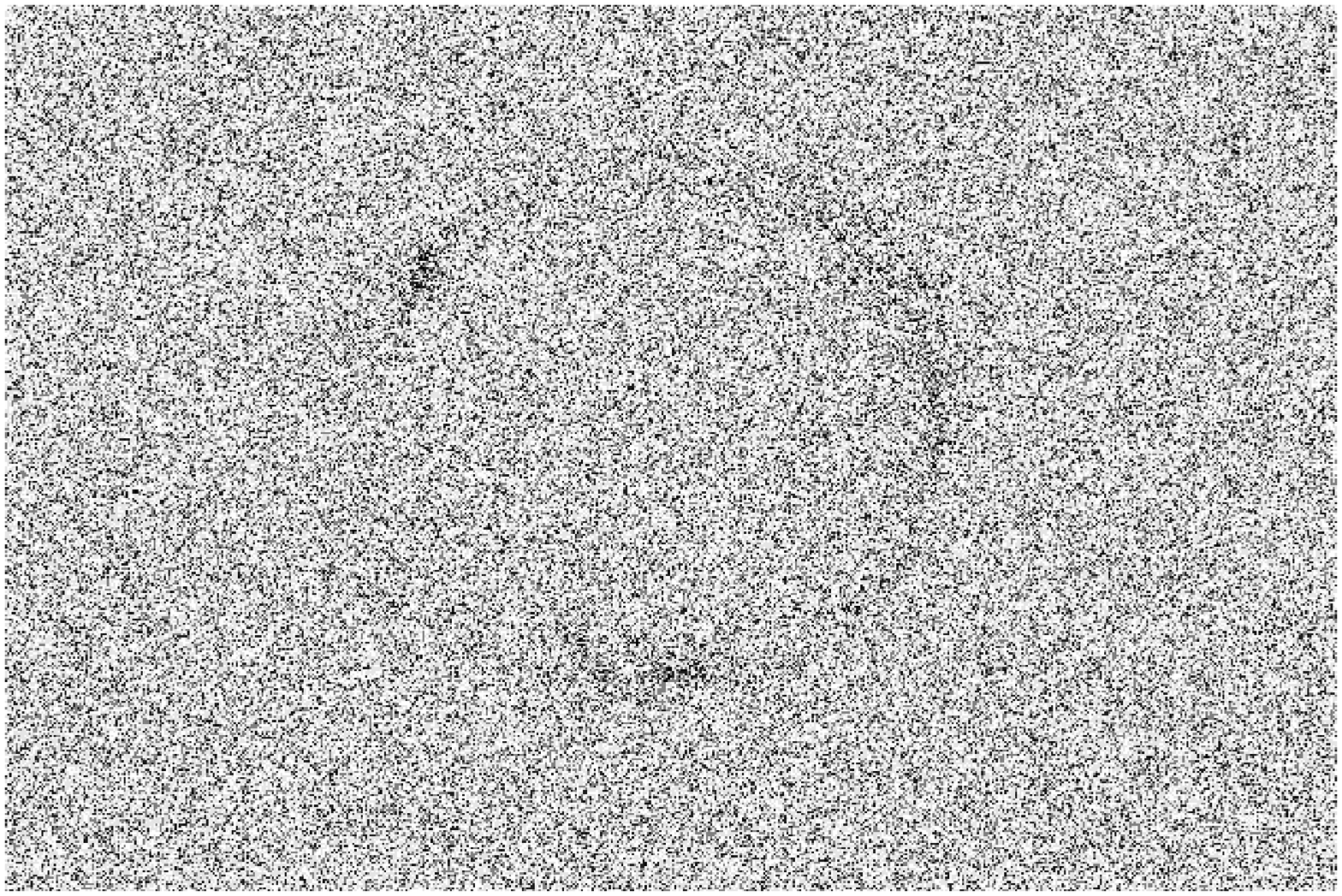}
\includegraphics[angle=0,scale=.24]{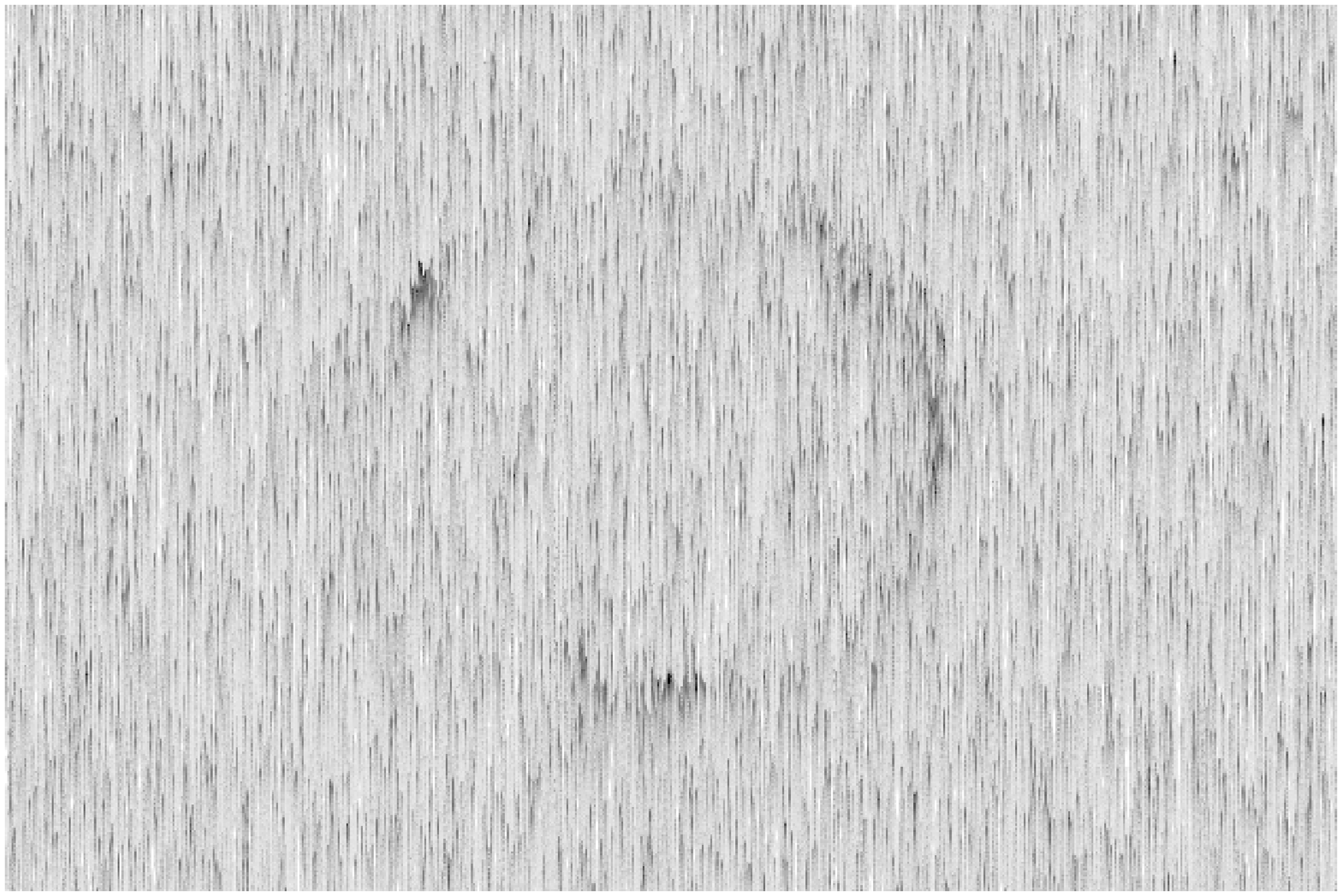}
\includegraphics[angle=0,scale=.24]{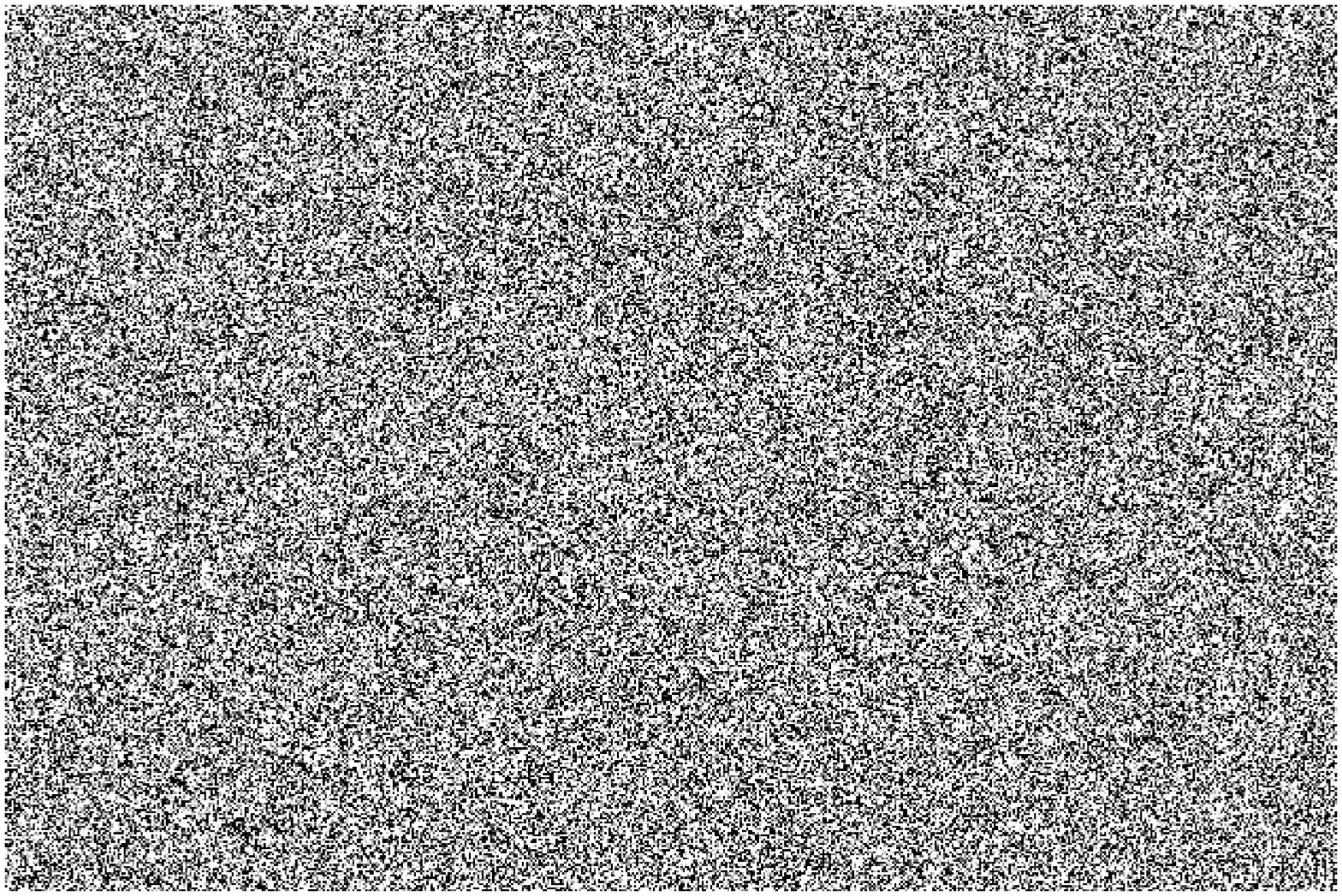}
\caption{This figure demonstrates how signal can be lost due to charge transfer inefficiency of the WFC3/UVIS CCD. Left: An assumed F275W (LyC) image based on the F606W image scaled down by 6 magnitudes with a typical background (dark+sky). Middle: The effect of CTI has been added based on the CTE Forward code provided by STScI. The effect of CTI is clear, as it has smeared the signal in the read-out (vertical) direction. Right: The simulated image is shown after adding read noise (which is not affected by CTE).} 
\label{fig:CTEforward}
\end{figure*}

\section{Method to account for CTI}
As mentioned above, we do not detect a significant LyC signal in the F275W image. However, it is possible that the depth of the image was significantly affected by CTI. If that is the case, we can calculate the threshold flux above which we should have detected the signal from the Horseshoe, including the effects of CTI. To account for CTI we should either be able to correct for it or simulate its effect on our assumed images. As the signal is essentially lost in our LyC filter, there is not enough flux to make a correction feasible. Therefore, we must forward-model the effects of CTI, to determine a flux threshold, above which we would have detected the Horseshoe despite CTI. For this analysis, we use the ``CTE forward'' code released by STScI \footnote{http://www.stsci.edu/hst/wfc3/tools/cte\_tools}, which simulates the effects of CTI on a provided image. 

\subsection{Finding the detection limit}

We first make an ``ideal'' image in the F275W filter, add the expected background, simulate the effect of CTE, and finally add the read-noise to produce an individual F275W exposure. We then add them to produce a simulated full-depth F275W image. Below we discuss details of each of these steps.  
 
\subsubsection{Assumed image}
\label{sec:aperture}
Because we do not know the morphology of the escaping LyC, we naively assume that the LyC has a similar morphology to that of the non-ionizing UV continuum in the F606W image. We then simply scale the flux from the F606W image down to simulate F275W images of various fluxes.  Because the F606W and F275W images were taken at different orientation angles (relative to the read-out directions), we also rotate the F606W image of the Horseshoe and place it in the same location as the expected position in the F275W image.

  \subsubsection{Background}
The CTI is very sensitive to the background level. Even a small background is effective at partially occupying the electron traps, resulting in reduced CTI. Therefore, it is imperative to carefully assess and add the background (sky + dark) electrons to our image before implementing the CTI effects.  
We calculate the dark current and sky backgrounds to be 1.53 $e^{-}$ pix$^{-1}$ exposure$^{-1}$ and 0.35 $e^{-}$ pix$^{-1}$ exposure$^{-1}$ respectively. Therefore, we assume a background with a Poisson distribution with an average of 1.88 $e^{-}$ pix$^{-1}$ exposure$^{-1}$. 
 
\subsubsection{Charge Transfer Phase and Read-noise}
To simulate the effect of CTE on our simulated images we ran the CTE-Forward code provide by STScI, assuming various total LyC fluxes. Afterward, the read noise was added to each pixel as a random number with a Gaussian distribution with a standard deviation of 3.0, consistent with the read noise for the four WFC3/UVIS amplifiers. This results in a simulated image which is a statistical representation of what would have been observed by the CCD, albeit without accounting for cosmic rays. 
 
 \subsubsection{Stacking the images}
We simulated 20 of these F275W images and stacked them to obtain a full-depth stacked simulated image. Now we are able to investigate whether, for an assumed LyC flux, the Horseshoe would be detected with $3 \sigma$ confidence in our three different apertures. 

Of course, our ``detection level" derived above is noisy, because it relies only on a single simulated stacked image, and the signal in this image will, by definition, vary by the input standard deviation multiplied by the square root of the number of pixels. Therefore, to more accurately assess the expected detection threshold, we produce five instances of these simulated images at the same LyC flux level and use the average of their detection levels. We had to using a higher $S/N$ background, because the Poisson variations in the background are large compared to the background and such variations can affect the level of CTI in each pixel.

\section{Results and Discussion}
\subsection{Observed Flux Ratios}

\begin{figure}
\includegraphics[angle=0,trim=1cm 0cm 0.5cm 0.5cm,width=1.02\columnwidth]{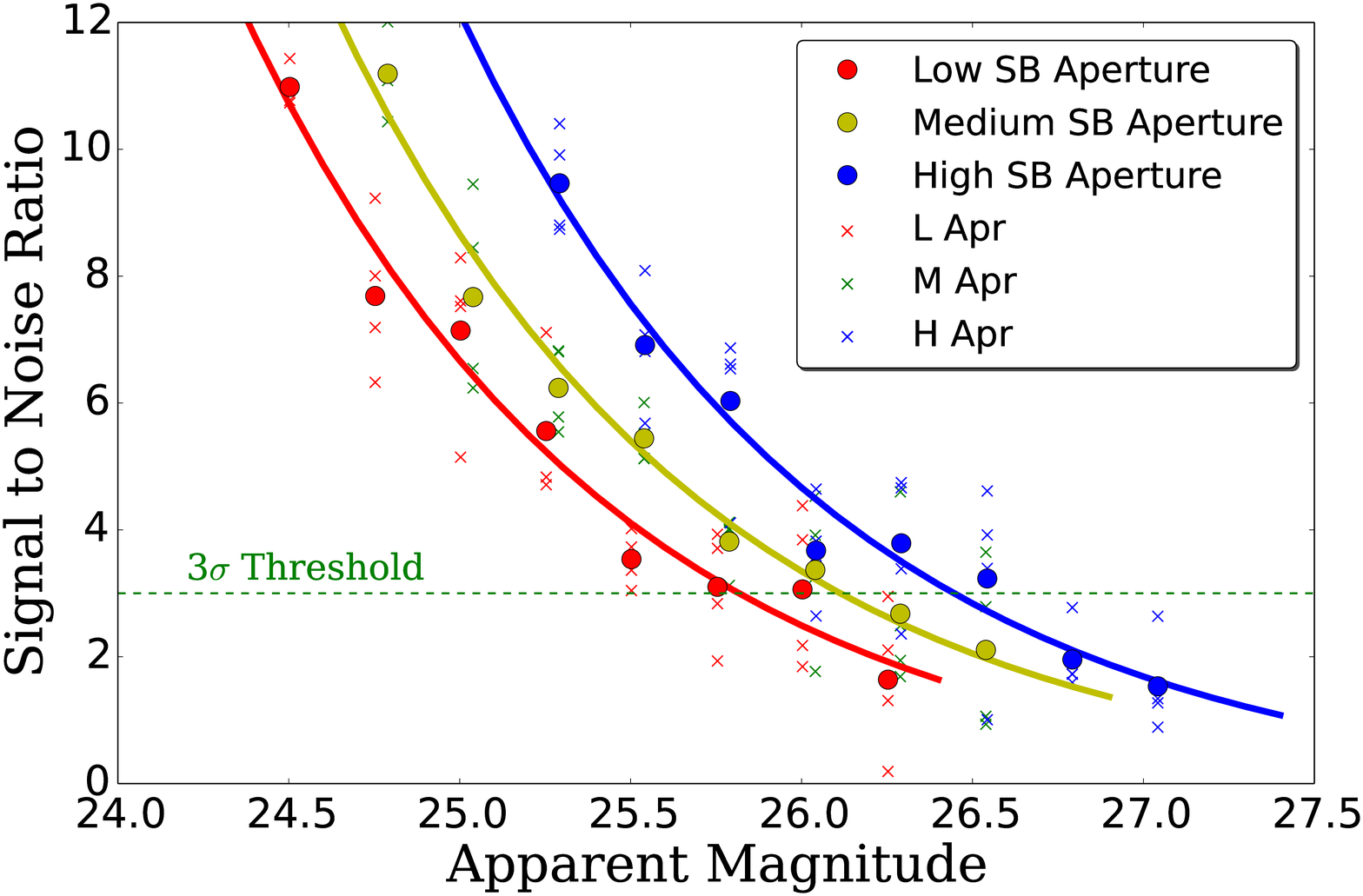}
\caption{ The measured signal to noise ratios of simulated images ($\times$ symbols) as a function of input F275W magnitudes, for the three apertures L (red), M (green) and H (blue). For each of the input magnitudes the average detection levels are shown  with filled circles in the corresponding color. For each aperture we fit a curve (solid lines) to the average values to determine at what magnitude the signal would be detected at 3 $\sigma$. These ``threshold magnitudes'' for each aperture are also listed in Table~\ref{tab:apertures}.}
\label{fig:result}
\end{figure}

We generated simulated images with eight different input magnitudes, spanning a range of two magnitudes with steps of 0.25 magnitude (close to where we expect to have $3 \sigma$ detection within our apertures). For each of the stacked simulated images, after removing the background, we measured the level of signal and noise within each aperture. Fig.~\ref{fig:result} shows the measured $S/N$ of each of the images in the three apertures, as a function of the input magnitude within that aperture. For each of the input magnitudes, the average $S/N$ is then plotted as filled circles. To find the input magnitude that results in a $3 \sigma$ detection, we fit a curve to the filled circles, assuming that input flux is linearly proportional to the detection level, and determine at what magnitude the significance of the detection would be greater than $S/N>3$. These threshold magnitudes and the corresponding F606W magnitudes are listed in Table \ref{tab:apertures}. All magnitudes have been corrected for Galactic extinction at the location of the Horseshoe based on \citet[][]{Schlafly11} ($A_{F275W}=0.258$ mags and $A_{F606W}=0.117$ mags). 

For comparison, we have also listed in Table \ref{tab:apertures} the escape fraction limits based solely on the background noise and not considering signal loss due to CTI. Interestingly, the limits are very similar, meaning that the CTI losses are not large for our measurements. The CTI results in 0\%, 5\%, and 17\% increase in the limiting flux respectively for L, M, and H apertures. Not surprisingly, the CTI effect on photometry is not significant in the largest apertures, as the typical trapped e$^-$ is released $<10$ pixels behind the original pixel \citep[][]{Rafelski15}. Therefore, CTI losses in the photometry only become a concern when the aperture width (in the readout direction) approach this length scale.

\begin{deluxetable}{lccc}[b]
\tablecaption{The characteristics and measured quantities in each aperture}
\tablehead{\colhead{Apertures} & \colhead{\bf L} & \colhead{\bf M} & \colhead{\bf H}}
\startdata
$\sigma$ above background in F606W \tablenotemark{a}      & 1.3 & 3.0 & 5.0\\
number of pixels            & 14302 & 7514 & 3278\\
magnitude in F606W           & 19.83 & 20.12 & 20.62\\
detection level in F275W                  & -1.23 & -0.99  & 1.08 \\
3 $\sigma$ limiting magnitude in F275W image \tablenotemark{b}     & 25.82 & 26.17 & 26.62\\
$f_{esc,rel}$ for limiting magnitude &  0.079 & 0.075 & 0.078 \\
3 $\sigma$ threshhold magnitude in F275W \tablenotemark{c} & 25.82 & 26.11 & 26.45\\
$f_{esc,rel}$ for threshold magnitude   & 0.079 & 0.079 & 0.092 
\enddata
\label{tab:apertures}
\tablenotetext{a}{The selection criteria to choose the apertures}
\tablenotetext{b}{Based on the depth of the image}
\tablenotetext{c}{Based on Fig.~\ref{fig:result} }
\end{deluxetable}

\subsection{Escape Fraction}
\label{sec:escape}
With the F275W $3\sigma$ threshold magnitudes and the F606W magnitudes in the corresponding apertures, we can calculate the limits on the non-ionizing to ionizing flux ratio and, ultimately, the relative escape fraction, using Eq. \ref{eq:fesc}.  In Table \ref{tab:apertures}, we list the $3\sigma$ limits on the relative escape fraction in all three apertures.

After including the effects of CTI, the upper limits on the relative escape fraction in each aperture are similar:  0.079, 0.079, and 0.092 in the L, M and H apertures, respectively, as the significant reduction in F275W noise in smaller apertures is offset by reduced F606W flux. Using a Calzetti extinction curve \citep{Calzetti97} and $E_{(B-V)}=0.15$ \citep[][]{Quider09}, we convert $f_{esc,rel} < 0.08$ to $f_{esc,abs} < 0.02$, well below the average value needed to maintain ionization at $z\sim7$ ($f_{esc,abs}\sim0.2$) \citep[e.g.,][]{Robertson15, Bouwens15b}.

\begin{figure}[b]
\includegraphics[angle=0,width=\columnwidth]{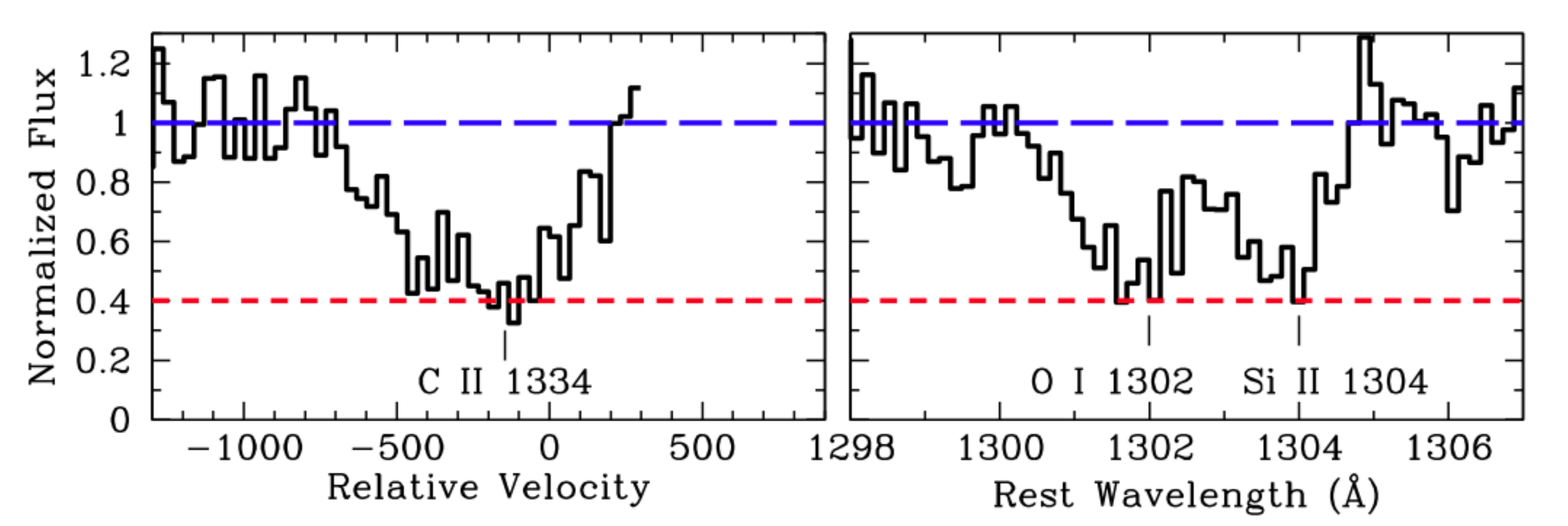}
\caption{Three low ionization absorption line profiles from the high resolution ($R\sim4000$) spectrum of the Horseshoe \citep[][]{Quider09}. Though these ions have different abundances, and the transitions have different oscillator strengths, the depth of each feature is the same, $\sim0.4$ of the continuum flux density, suggesting a covering fraction of $\sim0.6$ for the low-ionization gas. As this low-ionization gas is also where neutral hydrogen would reside, this could imply that as much as 40\% of the UV-bright disk may be unobscured, allowing LyC photons to escape along those sight lines.}
\label{fig:lines}
\end{figure}

As mentioned in Section \ref{sec:intro}, the high resolution, rest-frame UV spectrum of the Horseshoe has multiple, resolved interstellar absorption lines of low ions with depths of $\sim60$\% of the continuum flux density \citep[][]{Quider09}. Some of these lines are shown in Fig.~\ref{fig:lines}. These lines have very different oscillator strengths but similar absorption depths, especially for the case of Si {\sc ii} transitions, which all originate from the same ion and the relative depths are therefore independent of the metallicity or ionization. \citet{Quider09} therefore conclude that the depth of the lines is not due to the column density of a foreground screen. Rather, the depth is dictated by the fraction of the UV-bright disk that is covered by clouds that are opaque in these lines. Hence for the Horseshoe, we expect a $\sim0.6$ ``covering fraction'' of low-ionization gas where neutral hydrogen would reside. Based on the picket fence model and the discussion in the Appendix, this in turn implies $f_{esc,rel}\sim0.4$.  

The reported $3\sigma$ limits in our apertures of $f_{esc,rel}<0.08$ are roughly five times lower than the expected escape fraction inferred from the depth of the low-ion absorption lines in the spectrum of the Horseshoe. One reason for this discrepancy could be the existence of a rather opaque IGM along the LoS to the Horseshoe.
To obtain our measured $f_{esc,rel}$ we adopt the average transmission of the IGM through the F275W filter ($e^{-\tau_{H_{I,IGM}}}=0.402$). But, as is evident in Figure \ref{fig:IGMT}, this value is not representative of the nearly bimodal distribution of the IGM transmission. Specifically, there is an almost 20\% chance that the Horseshoe lies along a line of sight that is at least $5\times$ more opaque than the average transmission assumed here (less than 0.08), which could thus explain our non-detection, even if $f_{esc,rel}$ is 0.4 as suggested by the depth of the low ionization interstellar absorption lines. On the other hand, the IGM transmission could also be significantly higher than the average value, in which case our $f_{esc,rel}$ limit would be even lower. 

There are still a number of additional possible causes for a discrepancy between the transmission in the low ions and the relative escape fraction. Many of these reasons have been mentioned by \citet{Jones13}, but we discuss them here in the context of the Horseshoe;

1.  The absorption depths of the low-ions give only a covering fraction of the {\it non-ionizing} UV disk (from $\sim 1250-1700$ \AA\ ). However, because the LyC-emitting regions are short-lived, they likely comprise only a subset of the area that is emitting non-ionizing UV light. Therefore, a clear line of sight toward a non-ionizing UV-emitting region does not necessarily imply that LyC photons will escape. Furthermore, it is likely that the LyC-emitting regions have {\it higher} columns of dust and gas toward them, as they are younger and more embedded in their birth clouds. Indeed, this is the reason that extinction of H {\sc ii} regions is typically higher than extinction of other regions at the same wavelengths \citep[][]{Calzetti97}.  Therefore, the transmitted emission in the low-ion absorption lines should be considered an upper-limit to the possible LyC escape fraction. 

2. The LyC absorption by hydrogen acts at all wavelengths below the Lyman limit, whereas the absorption lines absorb at one specific wavelength. Therefore, one must be careful in interpreting the absorption lines. The velocity structure of the absorbing gas is critical when converting the depth of the absorption lines to covering fractions.  For example, if half of the galaxy is covered by gas outflowing at 0-100 km s$^{-1}$, and the other half is covered by gas outflowing at 100-200 km s$^{-1}$, then the depth of the absorption line (in a high-resolution spectrum) will never be more than 50\% of the continuum.  However, the LyC in such a scenario would be 100\% absorbed because it is insensitive to the velocity of the outflow. Thus, once again, the LyC escape fraction should generally be lower than expected given the absorption line depth. 

3.  If there are foreground neutral clumps with low velocity dispersion that are not resolved in wavelength, the profile of the absorption line will be smoothed, and the measured transmission will reflect only an upper limit on $f_{esc,rel}$. In our case the Keck/Echellette Spectrograph and Imager (ESI) spectra of the Horseshoe has the resolution of $\sim 75$ km s$^{-1}$. Though the velocity width of the absorption profiles is much larger than this value, it is still possible that there exists unresolved narrow components.

4. These absorption lines are due to resonance transitions from the ground state. Therefore, the photon can be scattered and re-fill the absorption line if the slit encompasses a large fraction of the scattering gas cloud \citep{prochaska11, rubin11, scarlata15}. Thus, the covering fraction may be larger than implied by the depth of the absorption lines, and the relative escape fraction will be lower.  

5.  If the LyC absorbing gas is very low metallicity \citep[][]{Fumagalli11} perhaps because it is inflowing from the IGM, then the metal line absorption will not be strong, but the hydrogen opacity will still be large and can absorb the LyC.  Though, given the significant amount of enriched outlawing material, and the large accumulated stellar mass, it seems unlikely that a significant fraction of the absorbing gas would be very low metallicity. However, in such a scenario, the absorption line transmission should be treated as an upper limit. 

Because of the reasons outlined above, we believe a non-uniform coverage of low-ionization metals is a necessary, but not sufficient, condition for significant escape of Lyman continuum.  Hence, any estimates of $f_{esc,rel}$ based on these absorption lines should be interpreted as upper-limits. 

Although $f_{esc,rel}$ may be significantly lower than predicted by the transmission in low-ion absorption lines, it is worth noting that the two values will be better correlated in more compact galaxies because many of the issues raised above that can cause the two to differ are mitigated significantly if the galaxy is extremely compact ($<100$ pc).  First, the ionizing and non-ionizing UV continua are likely emitted from the same regions (a single star-forming region).  In contrast, in a large galaxy, much of the non-ionizing flux is likely emitted from regions with no current star formation (and thus no LyC production).  Second, the smaller the galaxy, the likelier it is that the absorption of light from different parts of the galaxy is caused by the same absorbers, especially if the galaxy size approaches the typical size of an absorbing neutral clump in the ISM or CGM.  Thus, one has to worry less about clumps of different velocities covering different parts of the disk.  

Indeed, \citet[][]{Heckman11,Borthakur14,Alexandroff15} have been investigating the relative escape fraction, both directly and indirectly from luminous and extremely compact galaxies, which they refer to as dominant central objects (DCOs, ionizing UV sizes of $< 100$ pc).  Thus, it may be true that the absorption line transmission reasonably predicts the relative escape fraction in these galaxies, but it may not be the case in all galaxies. 

The intrinsic size of the Cosmic Horseshoe galaxy has been measured in both the non-ionizing UV and H$\alpha$ by \citet{Jones13A}.  In both cases, the emission is coming from an elongated region that is $\sim0.2$ kpc $\times0.4$ kpc, significantly smaller than the average UV size of galaxies of similar luminosity which typically have diameters of $\sim 3-4$ kpc \citep[][]{Law07}. With such a small area, we might expect that some of the issues above would be mitigated.  However, this area is still large enough that it likely consists of many distinct star-forming regions, and may still be covered by a range of clump distributions along the LoS.  To understand the efficacy of these indirect estimates of $f_{esc,rel}$, we must directly image the LyC of a large sample of galaxies with high transmission in the low-ion absorption lines.

\section{Summary and Conclusion}
In this study, we have attempted to measure the Lyman continuum escape fraction of the Cosmic Horseshoe, a highly magnified, star-forming galaxy at $z=2.38$. The high resolution rest-frame ultraviolet spectrum of the Horseshoe shows only $\sim60$\% absorption in the resolved interstellar absorption lines of low ions (e.g. O{\sc i}, C{\sc ii}, Si{\sc ii}), suggesting a patchy foreground neutral gas distribution \citep{Quider09} and a relative escape fraction of $f_{esc,rel}=0.4$. Given the high magnification, the well-suited redshift for the existing WFC3/UVIS filters, and the suggestion of a partially transparent foreground gas distribution, we obtained a 10 orbit image of the Lyman continuum (at $\sim800$\AA) with the WFC3/UVIS F275W filter.  

We made and subtracted enhanced darks that contain the structure seen in the actual darks.  We then forward modeled the effects of charge transfer inefficiency of the WFC3/UVIS CCDs to determine at what flux density we would no longer be able to detect the LyC. Because the photometric apertures are large (relative to unlensed galaxies), we find that the effects of charge transfer inefficiency on our photometry are quite small (17\% affect to the LyC photometry in the smallest [worst case] aperture). 

We do not detect significant LyC flux from the Cosmic Horseshoe, and the flux density limit gives an upper limit on the relative escape fraction of $f_{esc,rel}<0.08$ ($3\sigma$) when assuming average transmission through the IGM.  The upper limit is a factor of five lower than the value suggested by the significant transmission in the low-ion interstellar absorption lines. This suggests that the transmission in the interstellar absorption lines may not be a reliable proxy for the relative escape fraction \cite[though cf.][]{Borthakur14}. We outline a number of reasons why the transmission in the absorption lines of low ions may only give an upper limit to the escape fraction. Finally, we note that there is a 20\% chance that the transmission of the IGM along the line of sight to the Horseshoe may be five times lower than the assumed average, which would fully explain our non-detection even if the relative escape fraction were 0.4.  A study of a much larger sample of star-forming galaxies can statistically overcome the uncertain IGM transmission, and would definitively test the indirect method of measuring the escape fraction via the depths of the interstellar absorption lines.

\acknowledgments
The authors wish to recognize and acknowledge the very significant cultural role and reverence that the summit of Mauna Kea has always had within the indigenous Hawaiian community.  We are most fortunate to have the opportunity to conduct observations from this mountain.
Facilities: \facility{{\it HST} (WFC3,UVIS)},\facility{{\it Keck} (ESI)}.

\appendix

\section{Appendix A:  Covering Fraction to Escape Fraction}
\begin{figure*}
\centering
\includegraphics[angle=0,scale=.32]{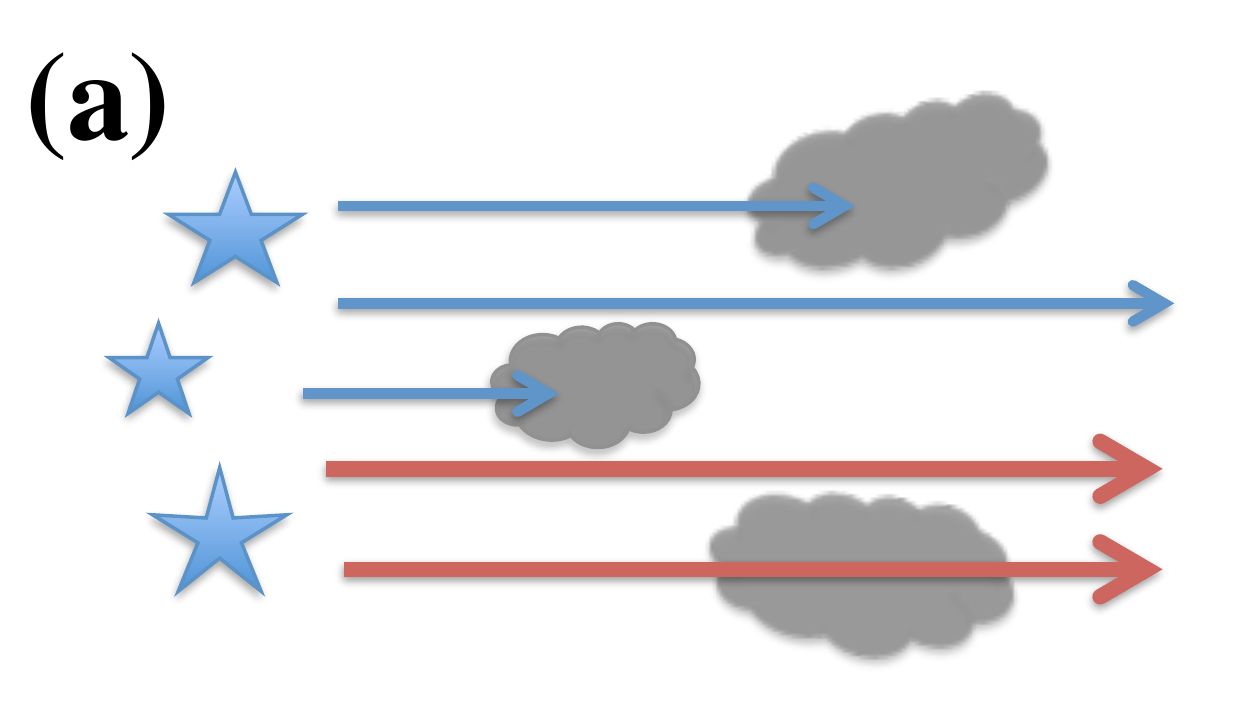}
\includegraphics[angle=0,scale=.32]{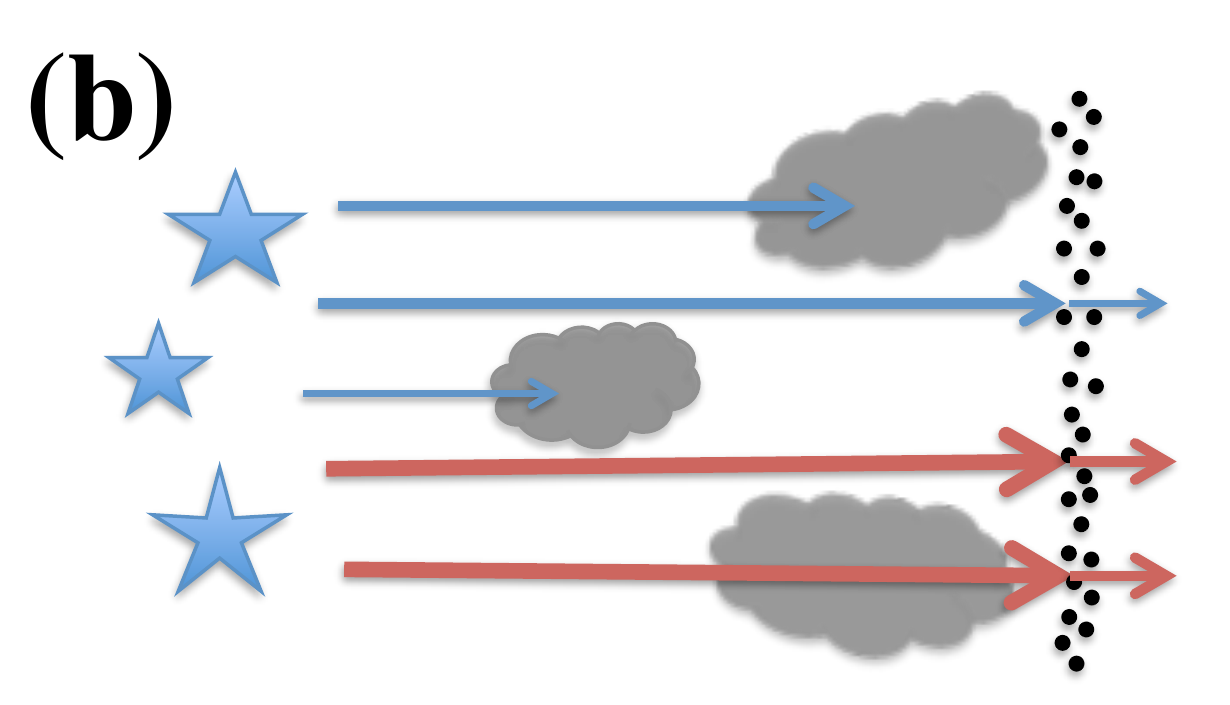}
\includegraphics[angle=0,scale=.32]{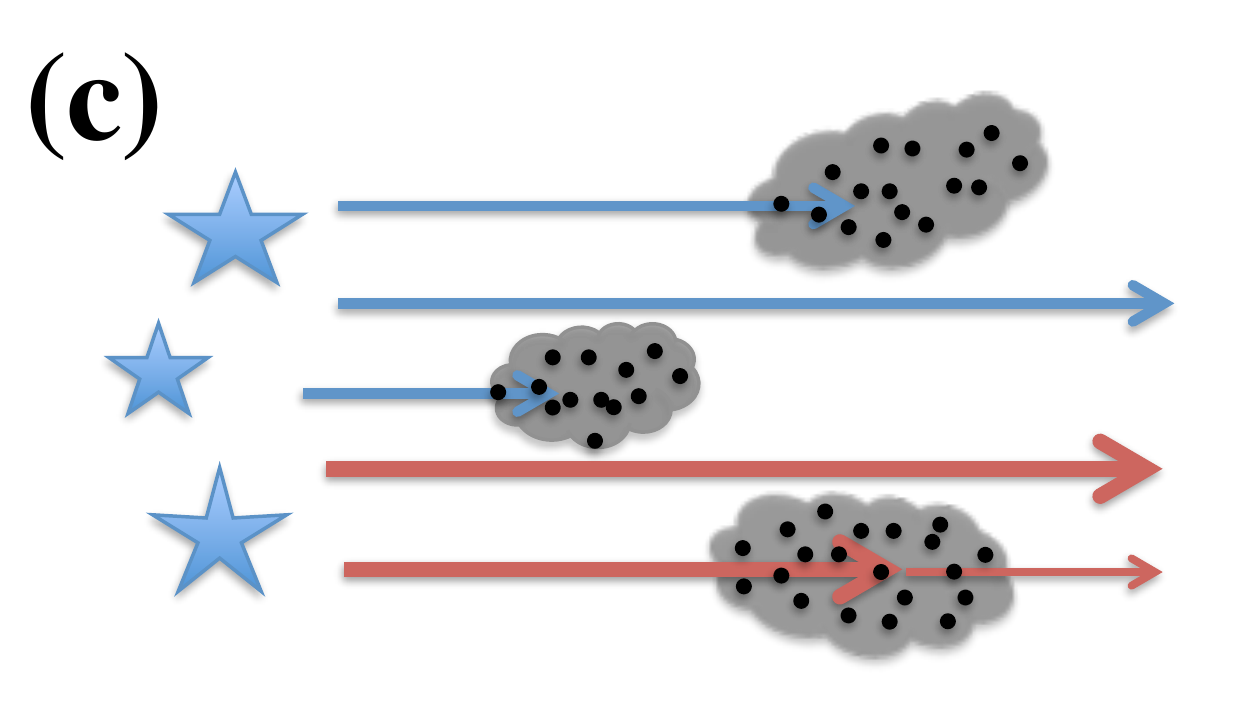}
\caption{The three simple models for the distribution of dust within a patchy ISM/CGM arm depicted: (a) no dust; (b) a uniform dust screen, (c) dust only within the gas clumps. The ionizing and non-ionizing fluxes are represented by blue and red arrows, respectively. } 
\label{fig:Models}
\end{figure*}

There are a number of definitions of the LyC escape fraction in the literature, ``relative escape fraction'', ``absolute escape fraction'', ``dust free escape fraction'', etc..  This has lead to some confusion about how they are related to the physical covering fraction or the depth of the low-ionizing absorption lines. To avoid this confusion, we derive here what exactly each of these terms represents for three simplified scenarios of dust distribution within a patchy interstellar/circumgalactic medium. These simplified cases are a) a dust free model; b) a uniform screen of dust and c) dust only located in dense clouds. Cartoons of these models are shown in Figure~\ref{fig:Models} where dusty regions are shown as solid black dots. In all cases, the gas is assumed to be of sufficient column density to be optically thick in the common low-ionization interstellar lines as well as the Lyman continuum.  We note that the precise location of the uniform dust screen in model (b) does not affect the calculations below. 

We also note that the covering fraction inferred by the depth of the absorption lines (denoted here by $C_{F'}$) is not necessarily the same as the physical covering fraction of the dense clouds ($C_{F}$). Observationally, we can only measure $C_{F'}$ which is defined as the ratio of the observed flux density at the wavelength of the absorption line (assumed to be completely saturated in the dense clouds), to the observed continuum flux density (at around 1500\AA):
$1-C_{F'}=\frac{F_{line,obs}}{F{cont,obs}}$.  Therefore, we report our escape fractions as a function of $C_{F'}$:

\subsection{Case (a) No dust :}  
For the dust free case, the LyC flux density and the flux density in the absorption lines get completely absorbed by the clouds while the 1500 \AA\ continuum flux density remains unaffected by the gas clouds:
 
 \begin{equation} 
{F_{LyC,out}=F_{LyC,stel}\times (1-C_{F})};   {F_{1500,out}=F_{1500,stel}}
\end {equation}
\begin{equation}
 {1-C_{F'}=\frac{F_{line,obs}}{F_{cont,obs}}=\frac{F_{1500,stel}\times(1-C_{F})}{F_{1500,stel}}=1-C_{F}}
\end {equation}

So here $C_{F} = C_{F'}$, which results in: 

\begin {equation}
{f_{esc,rel}=\frac{{\left( F_{LyC}/F_{1500}\right)}_{out}}{{\left( F_{LyC}/F_{1500}\right)}_{stel}}}=1-C_{F}=1-C_{F'}
\end {equation}

\begin {equation}
{f_{esc,abs}=\frac{F_{LyC,out}}{F_{LyC,stel}}=1-C_{F}=1-C_{F'}}
\end {equation}
In this case, $f_{esc,rel}$ and $f_{esc,abs}$ are the same. In the literature, this is sometimes referred to as the dust free escape fraction.

 \subsection{Case (b) Screening dust :}  
In this case we add a layer of uniform dust to the previous geometry. Now all of the flux densities are the dust attenuated flux densities in case (a):

 \begin{equation} 
{F_{LyC,out}=F_{LyC,stel}\times (1-C_{F})\times e^{-\tau_{dust,LyC}}};   {F_{1500,out}=F_{1500,stel}\times e^{-\tau_{dust,LyC}}}
\end {equation}
\begin{equation}
 {1-C_{F'}=\frac{F_{line,obs}}{F_{cont,obs}}=\frac{F_{1500,stel}\times(1-C_{F})}{F_{1500,stel}}=1-C_{F}}
\end {equation}

Again in this geometry we have $C_{F} = C_{F'}$, which results in: 

\begin {equation}
{f_{esc,rel}=\frac{{\left( F_{LyC}/F_{1500}\right)}_{out}}{{\left( F_{LyC}/F_{1500}\right)}_{stel}}}=1-C_{F'}\times e^{-(\tau_{dust,LyC}-\tau_{dust,1500})}
\end {equation}

\begin {equation}
{f_{esc,abs}=\frac{F_{LyC,out}}{F_{LyC,stel}}=1-C_{F'}\times e^{-\tau_{dust,LyC}}}
\end {equation}

Here, neither $f_{esc,rel}$ nor $f_{esc,abs}$ are not equivalent to the $1-C_{F'}$. We note that this is the geometry adopted by \citet{Borthakur14}. 

 \subsection{Case (c) Dust in clouds:}  
 In this case the dust closely traces the dense gas. We believe that this is the model that most closely resembles reality, as any significant dust will be accompanied by opaque columns of gas \citep[both in the LyC and the interstellar absorption lines, ][]{Gnedin08}. Though the dust has practically no effect on the line or LyC flux (they get absorbed predominantly by the gas), it attenuates the observed continuum flux. So we have:

 \begin{equation} 
{F_{LyC,out}=F_{LyC,stel}\times (1-C_{F})};   {F_{1500,out}=F_{1500,stel}\times(1-C_{F}+C_{F}\times e^{-\tau_{dust,1500}})}
\end {equation}
\begin{equation}
 {1-C_{F'}=\frac{F_{line,obs}}{F_{cont,obs}}=\frac{F_{1500,stel}\times(1-C_{F})}{F_{1500,stel}\times(1-C_{F}+C_{F}\times e^{-\tau_{dust,1500}})}\ne 1-C_{F}}
\end {equation}

Interestingly, in such a scenario the covering fraction based on the depth of the low ionization absorption lines is actually different from the physical covering fraction, i.e. $C_{F} \ne C_{F'}$.  As such we end up with the following relations for the escape fractions: 

\begin {equation}
{f_{esc,rel}=\frac{{\left( F_{LyC}/F_{1500}\right)}_{out}}{{\left( F_{LyC}/F_{1500}\right)}_{stel}}}=1-C_{F'} 
\end {equation}

\begin {equation}
{f_{esc,abs}=\frac{F_{LyC,out}}{F_{LyC,stel}}=1-C_{F}=1-\frac{C_{F'}}{C_{F'}+(1-C_{F'})\times e^{-\tau_{dust,1500}}}}
\end {equation}

Therefore, $f_{esc,rel}$ is, in fact, equal to the ratio of the observed flux density in the line to the observed continuum flux density, $f_{esc,rel}=1-C_{F'}$.  Since $1-C_{F'}$ is a common observable, it is best to compare this measurement to that of $f_{esc,rel}$ as we have done in this work.

\bibliographystyle{apj}
\bibliography{KavehRefs}

\clearpage

\begin{turnpage}
\end{turnpage}

\end{document}